# Implantable CMOS probes for high resolution Electrical Imaging of Local Field Potentials Across the Rat Barrel Cortex in vivo

Claudia Cecchetto[1,2,3], Mufti Mahmud[4], Vincenzo Sorrenti[5], Marta Maschietto[1], Enrico Chiarello[1], Mathias Schulz[6], Roland Thewes[6] & Stefano Vassanelli[1,3]

[1] Department of Biomedical Sciences, University of Padova, via Ugo Bassi 58B, 35131 Padova, Italy
[2] Optical Neuroimaging Unit, Okinawa Institute of Science and Technology Graduate University, OIST, Okinawa, Japan.
[3] Padua Neuroscience Center, University of Padova, via Orus 2/B, 35131 Padova, Italy
[4] Nottingham Trent University, United Kingdom
[5] Department of Pharmaceutical & Pharmacological Sciences, University of Padova, 35131 Padova, Italy.
[6] Faculty of Electrical Engineering & Computer Science, TU Berlin, Berlin, Germany

**PRELIMINARY DRAFT TO BE SUMITTED FOR THE PUBLICATION ON NATURE MATERIALS**

**Abstract**
High-resolution recordings of extracellular potentials are fundamental for the study of neuronal networks, the basis of neuronal coding and transmission. Here we present an innovative method for an in vivo electrical imaging of Local Field Potentials using novel implantable neural interfaces with a high-density array of 256 recording sites and a spatial resolution of 15 µm. Thanks to this technology, we analyze the propagation of whisker-evoked activity in a single barrel-column of the rat somatosensory cortex.

**Introduction**

Electrical recordings proved to be an optimal technique to investigate neuronal networks and complex functions as sensory processing, memory and behavior (Buzsaki, 2004; Obien et al, 2015; Hong et al, 2019). In the last 20 years, many advances were made in electrode production, starting from few channels to thousands of sensors embedded on the same probe. The technology behind probe manufacturing is also broad: metal electrodes (Tambaro et al, 2021; Lepaurulo et al, 2020; Steinmetz et al, 2018) as well as CMOS-based chips (Cecchetto et al, 2015, Mahmud et al, 2020) are commonly used in research. However, some probes are specifically designed for basic research and murine models (Jun et al, 2017; Steinemetz et al, 2021), thus are difficult to adapt for clinical practice, e.g. for brain implants and deep brain stimulation. To be eligible for medical applications, the ideal probe must have a highly translational capacity/configuration, limited dimensions, high biocompatibility and stable properties for long term implants.
CMOS based probes are promising candidates for clinical use, because of their thin layer of TiO2, which is a highly biocompatible material and can establish a tight capacitive coupling with the neural tissue, that ensure stable and reliable recordings.

The rat barrel cortex is a well-known example of topographic mapping, where each barrel encodes the tactile information coming from a particular whisker in the mystacial pad of the animal (Land & Simons, 1985; Woolsey & Van der Loos, 1970). A pathway comprising just three synapses connects the primary afferents carrying information from the whisker follicle receptors to the final link into the cortex, passing through the trigeminal ganglion (TG), the brainstem and the thalamic nuclei (VPM, Ventral Posterior Medial nucleus and POm, Posterior Medial nucleus). The larger barrels (so called because of their characteristic curved shape that resembles a barrel stave) are about 500 µm in width and about 300 µm along the axis of the row, separated by 70 µm wide septa (Alloway, 2008). The barrel-column, which forms the cortical processing unit of a single vibrissa, is composed of a vertical array of cells running orthogonally to the six-layered structure of the cortex, below and above the corresponding neurons of the single barrel in layer IV. Each column measures about 300 µm in width and runs through the entire depth of the cortex, about 2 mm long in rats. Each barrel-column is composed of an archetypical circuit that is repeated in each column, with each large barrel containing approximately 2000 neurons (Meyer et al, 2010): of these, about 75% are excitatory and 25% are inhibitory (Ren et al, 1992). Layer IV is the only one among the cortical layers that sends strong functional excitatory connections to all the other layers within the column (Staiger et al, 2000).
Instead, only layer V receives functional excitatory inputs from all the other cortical layers within the column (Staiger et al, 2000, Feldmeyer, 2012). In the sensory cortex, these columns form a spatial pattern similar to the arrangement of the whiskers on the mystacial pad. This distinctive barrel-shaped pattern can be easily seen from a horizontal section of the tissue obtained from layer IV.
Because of this simple grid-like organization, each barrel can be identified by its arc position (1-7, caudal to rostral) within a specific row (A-E, dorsal to ventral). Each barrel is selectively activated by the corresponding whisker on the contralateral mystacial pad (Simons, 1978; Welker, 1976), in a way that the topological position of the barrel within the barrel field is identical to the topological position of its corresponding whisker within the muzzle.
Thanks to its unique structural and functional organization, the barrel cortex has served as a test-bed system for several new methodologies, facilitating the study of the neuronal mechanisms at the base of sensory coding (Fox, 2008).
Neuronal population activity in the mammalian cortex is known to display a wide range of activation patterns. Both spontaneous and sensory-evoked activity typically shows propagating behaviors, manifesting as travelling waves (Wanger et al, 2013) but up to now there are no studies directed towards a high-resolution electrical mapping of these waves in the rat barrel cortex in vivo. Here, a novel typology of CMOS (Complementary Metal-Oxide Semiconductor)-based neural probes for in vivo high-resolution recording of Local Field Potentials (LFPs) is presented, featuring a high-density array of 256 recording sites with 15 μm spatial resolution.
The last layer deposited on the surface of the CMOS probe is a 30 nm thick $TiO_2$ layer (Schroder et al, BioCAS, 2015; Schroder et al, IWASI, 2015; Schroder PhD thesis, 2016). This thin layer establishes a strong capacitive coupling with the cellular membranes of neurons and offers excellent biocompatibility with the cortical tissue. Thanks to these properties, we envisage a high translational capacity of these neural probes, which cannot be solely used for basic research in neuroscience but also for chronic implants, ideally in patients affected by degenerative diseases, as already done for deep braining stimulation (DBS) (Little et al, 2013; Ashouri et al, 2018). The first cortical profiles of the whisker-evoked responses obtained with this technology are shown and discussed.

**Results**

Needle chips could be used multiple times and for long experiments: as summarized in Table 1, up to 20 experiments could be performed with the same probe, resulting in a total time of implantation longer than 40 hours. A single recording session from a rat could be run up to 4 hours, where electrical performances remained stable and no tissutal debris can be detected on the surface of the needle after a careful cleaning protocol at the end of each experiment (see 'Cleaning procedures' section).
Voltage operation points set at the acquisition interface were kept constant, even after several experiments. Nevertheless, the average calibration factor (see 'Methods') for probe #5 decreased of about 25% of the initial value after 20 experiments. This change though could be minimized by the calibration procedure performed for each experiment, where the calibration factor is computed at the beginning of every recording session.

**Discussion**

The depth profile obtained by averaging and concatenating single traces recorded at the various depths (see Methods and Figure 4) resembled the ones previously reported in literature. Specifically, the shape of the evoked peaks triggered by whisker deflection was very similar to those recorded by Roy et al, 2011 by means of a 16-channels 100 μm pitch silicon probe from Neuronexus. The overall behavior of the LFP traces in the central layers (from layer II to layer V) was also comparable to those recorded by Einevoll through a linear multielectrode array with 23 contacts spaced at 100 μm. Small differences in the shape of the LFPs recorded in the first superficial depths could reflect different electrical boundary conditions at the cortical surface in the used recording setups (as discussed also in Einevoll et al, 2007). Notably, the cortical profiles reported here have a much higher resolution, being the pitch of these probes almost 10× smaller of those used in other works. Thanks to these high-density sensors the intra-barrel and intra-layer propagation of the sensory-evoked signals could be investigated. At each time instant, the electrical potentials probed by the 256 sensors on the tip of the array were color-mapped in a 16×16 image where each recording site was represented by a specific pixel. Figure 3 reports five 16x16 LFP frames, sampled at five distinctive time instants of the evoked response (labelled on the top of the frame and marked as circles on the time-amplitude plot on the left). At every cortical depth, the video of the acquired LFPs was obtained by concatenating all the frames along the full time course of the signal (Supplementary Video 1). The reconstructed video could thus be used to follow the local propagation of the signals in a cortical region that is 225 × 225 µm$^2$ wide. The lateral extent of this region was shorter than the average diameter of a single barrel (about 500 µm for the larger ones, Alloway, 2008), while the height of the array covered almost the thickness of a single layer (Armstrong-James et al., 1992). By carefully analyzing the direction of propagation and the equipotential lines across the entire acquisition matrix of single traces acquired at a specific depth (see Figure 3 and Supplementary Video 1), it was possible to understand the sequence of cortical processing in a single barrel and reconstruct the architecture of intracolumnar neuronal projections, basing also on available anatomical data and previous electrophysiological recordings. For example, in the superficial layer I signals came from the bottom-right corner moving upwards, covering only the bottom-half of the array. Moving to the deeper layers, the portion of the array interested by this vertical flow going from the bottom to the top increases gradually, covering about 3/4 of the array (layer II/III, see Figure 3, top row, 400 µm) up to the whole matrix (layer IV). Layer IV showed a peculiar behavior, with a negative signal initially coming from restricted areas (about 3 × 3 sensors) and radially expanding to the whole array. In Figure 3, center row, 750 µm, the signal is coming from the top left corner (frame at 16.3 ms) and spreads quickly to the whole matrix (frame at 20.1 ms). This finding agreed with the known functional architecture of the barrel cortex, as this layer is the main target of thalamic axons (from VPM) and acts as a "hub" of intracolumnar information processing, distributing the signal coming from the whisker to all the other cortical layers in the same barrel-column (Feldmeyer, 2012). Layer Va presented the widest range of behaviors, as two different ways of propagation were found: one coming from the bottom and moving upwards and one coming from the left side moving downwards but usually returning back to the top left corner (like the one in Figure 3, bottom row, 1200 µm). In most of the traces activity from the overlying layer IV is evident and this broad spectrum of neuronal dynamics might be the result of the concurrent input/activity coming from layer IV and directly from the thalamus (POm). Activity in layer Vb showed a clearer behavior, propagating from the top-left corner to the right-bottom corner and back. In layer VI, the deepest one, the most intense activity was again confined in one half of the array, but this time the upper one. There, the flow of excitation moved from the top-left corner to the right-bottom corner and back. In Figure 3, the standard deviation (STD) of the signal is computed between the 256 electrodes for each depth and plotted on the left panel over the time-course of the average response. Here, the STD is not a measure of the noise of the signal, but rather emphasizes the non-uniformity of the spatial patterns of the signal over the 16x16 matrix in each time instant and gives a first estimate of the voltage gradients generated by the barrel microcircuitry during a single evoked response.

The cortical profile across the whole depth of the barrel column could be reconstructed by simply concatenating the 16×16 LFP signals recorded from successive depths. The full cortical profile could be sampled also through a single insertion of the 4x64 array in the tissue (Figure 4), with half of the spatial resolution (33 µm instead of 15 µm). The two datasets obtained with the two array configurations were aligned and compared (Figure 4 and Supplementary Video 2): signal amplitudes were in the same voltage range (-2.5 ÷ +1.0 mV) and, most importantly, the spatial extent of excitatory streams was very similar in the time instants considered along the time course of the response (onset, peak and repolarization). The interlayer propagation of the evoked response could be thus investigated: the negative signal arose from layer Va (Figure 4, 2$^{nd}$ column labelled with '15.2 ms', depth range: 960-1320 µm) and rapidly expanded to upper layers, gradually invading almost the whole column (Figure 4, 3$^{rd}$ column labelled with '16.8 ms'). Here, equipotential lines in layers IV-Va were concentric and centered around 1200 µm, while equipotential lines in the upper layers are almost horizontal: thus, the direction of the voltage gradient matched the vertical organization in cortical layers of a barrel. The evoked signal then faded away, lingering in layer IV (Figure 4, 5$^{th}$ column labelled with '23.2 ms', depth range: 600-840 µm) before disappearing completely from the recorded column. There, the predominant contribution to the initial part of the LFP signal was from layer V, apparently in contrast to the "hub" role of layer 4 in the intracolumnar information processing supported by connectionists (Feldmeyer, 2012). Nevertheless, the signal from layer IV spiny-stellate cells is usually overshadowed by that of layer V pyramidal populations, making the reconstructions of the anatomical origin of those signals more difficult. As a matter of fact, the LFP response evoked by the population of spatially compact layer IV spiny-stellate cells is much weaker than that evoked by the L2/3 and L5 populations of pyramidal cells in response to the synaptic input (Gratiy et al, 2011). The positive rebound of the signal after repolarization (t>40 ms from the stimulus) similarly invaded the whole column, with the strongest signals recorded from layer IV and Va (depth range: 600-960 µm, see Supplementary Video 2).

The time course of the neuronal activity generated in all the layers of the cortex by a single stimulation of a whisker could be visualized also in Figure 5. Here, the LFP signal in each row from 1 to 64 was obtained by averaging the signal acquired by the 4x64 array probe along the 4 columns of the corresponding row at every time frame. The average signal was then plotted on a 2D color map versus time. In Figure 5, the first 100 ms of the LFP signal following a single whisker deflection is shown, where row 1 corresponds to the cortical surface (0 µm) and row 64 corresponds to about 2016 µm deep in the cortex. Onset latencies and peak latencies from different cortical layers could be easily compared from this kind of plots: here, again, the sensory-generated negative peak first appeared in rows 32-48, corresponding to layers Va and Vb, having the shortest latency (12-15 ms). The dipole-like potential pattern between upper (rows 1-16) and deeper (row 23-56) layers is also highlighted in this plot. A long positive 'rebound' wave was then recorded between 30 to 70 ms in all the layers, mainly in rows 16-56, indicating an inhibitory epoch following the stimulus-evoked excitation of the neuronal network.

The position of the probe insertion site was verified using Nissl staining with Cresyl Violet acetate (see Supplementary Figure 1). The damage provoked by inserting the probe in the tissue is very limited and restricted to a diameter of less than 200 µm around the probe track (marked by the red dashed line). Being the lateral extension of the probe equal to 300 µm, we can consider this result very promising. Neurons around the insertion area do not seem affected or reorganized by the needle probe. Moreover, the track left by the neural probe confirms the orthogonality of the insertion into the barrel field region, as emphasized by the comparison with the anatomical atlas of the rat brain at 2.85 mm posterior from the bregma (adapted from Swanson atlas, 2008).

Tissue damage was assessed also through immunohistochemistry on 20 µm thick coronal slices using DAPI and IBA1 markers. Activated microglia, which is known to be a mediator of inflammatory pain, was found in close proximity to the insertion trace (see Supplementary Figure 2) in brains after about 3 hours of implantation. Nevertheless, the phenomenon was restricted to a small portion of tissue and the neuronal architecture (i.e. the position of somata) of the cortex did not seem to be impaired by the probe insertion, as revealed by DAPI immunohistochemistry (see Supplementary Figure 2).

**Methods**
*High-resolution neural probe.*
The needle-shaped probes are 10 mm long, 300 µm wide, 150 µm thick, and provide an array of 256 capacitively coupled recording sites arranged into a square 16×16 matrix with a pitch of 15 µm (Fig. X) or a rectangular 4x64 matrix with a pitch of 32 µm (see Figure 1 and Figure 2).
The last row of the array is located at 250 µm from the tip of the needle while the first row is at about 480 µm from the tip for the square matrix and at around 2270 µm for the rectangular one.

Needles are fabricated on the basis of a standard 180 nm CMOS process with 6 metal layers extended by specific post-processing steps to form thin (50 nm) Ti/TiN top electrodes covered by a 30 nm TiO2 dielectric on the chip surface (Schroder, BioCAS 2015). The standard CMOS process flow stops after deposition of the last $SiO_2$ intermetal dielectric covering the last CMOS metal (Al). The post-CMOS process starts with etching a top via hole which is filled with W, followed by deposition and patterning the top Ti/TiN electrodes with an octagonal shape at approximately 8 µm diameter. Then, the devices are diced out from the CMOS wafer using an inductively coupled plasma rapid ion etching process. Finally, the TiO2 dielectric is processed using an Atomic Layer Deposition (ALD) machine equipped with a specific handling tool enabling 3D deposition with very good uniformity on all chip surfaces. The thin layer of TiO2 (30 nm-thick) is deposited over the silicon surface so that the needle chip does not impair the neural tissue and vice versa, ensuring biostability and biocompatibility of the probe materials. An areal capacitance of 1.2 µF/cm$^2$ and a related relative dielectric constant of 40 are achieved at leakage currents below 0.1 µA/cm$^2$ within a suitable bias voltage window (difference of reference electrode potential applied to a potentiostat controlling the electrolyte on the chip surface and recording electrode potential within the chip) of 2 V width. The diced and fully processed needle chips are wire-bonded and glued onto a small printed circuit board (PCB) which provides a connector to interface electrically to the entire recording electronics.

*Data acquisition*

The 256 recording sites on the chips are electrically arranged as a 16 by 16 matrix, each row being externally connected to one of 16 independent, custom-made transimpedance amplifiers (TIA), which record and amplify the overall current from the rows with a bandpass freqyency range comprised between 4 Hz and 10 kHz. By applying a proper sequence of biasing voltages on the columns, the 16 recording transistors of each row are then time-multiplexed, by sequentially switching on one transistor at a time, while leaving the other ones off.

In the case of the 4x64 needle probes, each 'row' at the amplifier corresponds to a 4x4 subset of electrodes, where row 1 is the most distant group from the tip, while row 16 is the closest to the tip. Each 'column' of the amplifier corresponds to a single electrode of the selected 4x4 subset, starting from the top left corner and proceeding from left to right, top to bottom.

The multiplexed signals are then digitized by a commercial data acquisition (DAQ) card NI PXIe-6358 (National Instruments) up to 1.25MS/s at 16-bit resolution and saved to disk by a custom LabVIEW acquisition software.

The readout is performed sequentially row-wise, one full column being accessed simultaneously at the same time. After the last column has been accessed, all the recording sites (i.e. one full-frame) have been sampled, and the process repeats starting from the first column. One such full-frame corresponds to a single "time-point", however the signals recorded at different columns are sampled at slightly different times, giving rise to an imaging artifact analogous to the rolling shutter artifact common in CCD digital cameras. This must be accounted for in the data analysis (see following section).

This approach allows using a limited number of wires to access a much greater number of recording sites, at the cost of reducing the overall full frame-rate. For example, if the simultaneously operated differential analog inputs belonging to 16 recording sites of a column were sampled every 100 µs (column access time, see 'Implantation and recording procedures'), the whole array could be sampled in 1.6 ms.

Finally, these raw data were saved as multiplexed technical data management stream (tdms) files by a custom-made acquisition software built with LabVIEW (National Instruments, Austin, TX, USA).

*Surgical procedures*
Wistar rats were maintained in the Animal Research Facility of the Department of Biomedical Sciences (University of Padova, Italy) under standard environmental conditions. Animal-related procedures were performed in accordance with the European Union Guidelines for the Care and Use of Laboratory Animals and those of the Italian Ministry of Health (D.Lg. 26/2014). The study was approved by the Animal Care Committee (O.P.B.A.) of the University of Padua and the Italian Ministry of Health (Protocol number 447/2015-PR).
For all the experiments, P30-P40 rats were anesthetized with a mixture of Tiletamine (2 mg/100 g weight) and Xylazine (1.4 mg/100 g weight). The anesthesia level was monitored throughout the experiment by testing eye and hind-limb reflexes, respiration and checking the absence of whiskers' spontaneous movements. Additional doses of Tiletamine (0.5 mg/100 g weight) and Xylazine (0.5 mg/100 g weight) were provided every hour to maintain the anesthesia level constant. During the surgery and the recording session, animals were kept on a common stereotaxic frame, fixed by teeth and ear bars and observed with a stereomicroscope (Nikon SMZ800, Nikon Instruments Inc., Melville, NY, USA). The body temperature was constantly monitored with a rectal probe and maintained at about 37 °C using a heating pad. The fur of the head of the rat was trimmed and then, to expose the cortical area of interest, anterior-posterior opening in the skin was made along the medial line of the head, starting from the imaginary eyeline and ending at the neck. While the skin was kept apart using halsted-mosquito hemostats forceps, the connective tissue between skin and skull was gently removed by means of a bone scraper. Thus, the skull over the right hemisphere was drilled to open a window in correspondence of the somatosensory cortex, S1 (-1 to -4 AP, +4 to +8 LM) (Swanson, 2003). Meninges were then carefully cut by means of forceps at coordinates (-2.5 AP, +6 LM) to avoid the damage of the needle chip in the subsequent insertion.

All these procedures have already been described with the same details in [Mahmud M, Pasqualotto E, Bertoldo A, Girardi S, Maschietto M, Vassanelli S JNeurosci Methods. 2011 Mar 15;196(1):141-50].

A small bath of 37 °C warmed Krebs solution (in mM: 120 NaCl, 1.99 KCl, 25.56 NaHCO3, 136.09 KH2PO4, 2 CaCl2, 1.2 MgSO4, 11 glucose), that mimics natural brain fluids, was created on the head of the rat, where the reference ground electrode was immersed. The bath was maintained throughout all the experiment.

*Implantation and recording procedures*

The bath over the brain of the animal was grounded using a massive Ag/AgCl electrode (diameter of about 1 mm) connected to the main ground of the system, in a way that a differential signal can be obtained and the electrical modulation of the bath could be minimized.

The chip was moved, positioned and inserted in depth in the tissue by means of a Patchstar micromanipulator (Scientifica Ltd, East Sussex, UK) equipped with custom holders. The position of the probe in the cortical depth was first monitored through the microscope and then, after the impalement, through a dedicated software (Scientifica Linlab Control). By visual inspection, the tip of the chip was first put in contact with the cortical surface (defined as 0 μm depth) at the opening of the meninges and then lowered into the tissue perpendicular to the cortex, so that the first row of the array was positioned at 0 μm. For the 16x16 needle, each subsequent recording position required a 120 μm step vertical movement of the chip while for the 4x64 needle no additional movement through the cortex was needed.

In experiments performed with the 16x16 arrays, at each recording depth, the signal was first calibrated with a burst of 10 sinusoidal cycles (amplitude: 1 mV, f=20 Hz)} and then 60 evoked responses were acquired. The length of the acquired sweeps was 1.4 s and the intersweep interval was 1.3 s. The trigger for the stimulation of the whisker was set at 0.85 s from the beginning of each sweep. The global sampling frequency for multiplexed acquisitions of LFPs was set at 250 kHz, while the column access time was chosen to be of 100 μs. Therefore, the full-frame-rate, i.e. the final sampling frequency for a single sweep acquired by all the 256 channels, was equal to 625 Hz.

In experiments performed with the 4x64 arrays, a single careful insertion of the needle chip in the cortex (up to 2 mm deep) is needed. Signals are first calibrated with a burst of 100 synusoidal cycles having amplitude of 5 mV and frequency of 20 Hz, acquired in 6 s long sweeps. Then, 60 evoked responses were acquired in 10 s long sweeps with intersweep interval of 1 s. The trigger for the stimulation of the whisker was set at 0.15 s from the beginning of each sweep. The global sampling frequency was set at 1.25 MHz, while the column access time was chosen to be of 64 μs.

*Vibrissa Stimulation Protocol*

At the end of the surgery, contralateral whiskers were trimmed at about 10 mm from the mystacial pad. Single whiskers were deflected repeatedly by rapidly displacing a 25G hypodermic needle (BD Plastipak, Madrid, Spain) attached by means of a drop of superglue to a multilayer piezoelectric bender with integrated strain gauges (P-871.122, Physik Instrumente, Karlsruhe, Germany). The bender was driven by a power amplifier (E-650.00, Physik Instrumente, Karlsruhe, Germany) connected to a waveform generator (Agilent 33250A 80 MHz, Agilent Technologies Inc., Colorado, USA) through a closed-loop control able to monitor the actual movement of the piezoelectric bender by reading its strain gauges values. The bender was driven by providing square stimuli at 1 kHz, triggered by a custom LabView program (\url{www.ni.com/labview/}). At every cortical depth, single sweeps were recorded in response to these mechanical stimulations with a temporal delay of 2.7 s between subsequent deflections, in order to avoid any phenomenon related to adaptation.

*Cleaning procedures*
At the end of each experiment, the probe is carefully extracted from the brain by means of a micromanipulator and then extensively cleaned by applying a three steps protocol in an ultrasonic bath (SONOREX SUPER RK 514 BH), with each step lasting 15 minutes, in the following order: (1) deionized water, (2) 5% (gr/ml) sodium dodecyl sulfate (SDS; Sigma Aldrich) and (3) 5% (v/v) Tickopur R33 (Bandelin), the last two prewarmed at around 70 °C. The probe is finally rinsed in deionized water and checked under the microscope: if any tissutal debris is detected, the entire cleaning procedure is repeated until the microchip becomes perfectly clean.

*Signal processing, analysis, and visualization*
Raw signals sensed by different recording sites were saved as multiplexed technical data management stream (tdms) files by a custom-made acquisition software built with LabVIEW (National Instruments, Austin, TX, USA). The raw datafiles were processed in Matlab (MathWorks, Inc., Natick, MA, USA) using custom scripts. The data were converted to .mat format which were then processed, analyzed, and visualized as frames and discrete time videos.

*Data conversion*
The raw data in tdms files were imported to Matlab through custom scripts using TDMS C DLL interface provided by the National Instruments (http://www.ni.com/example/30957/en/). They first need to be down-sampled and de-multiplexed in order to recover the signals from each individual recording site.
One raw datafile consists of 16 waveforms (one for each row) which contain the actual signals from the recording sites, as well as transients due to the cyclic switching of the transistors during the multiplexing.
Each multiplexing cycle contains a transient, which is discarded, followed by a steady signal, whose data samples are averaged to obtain the instantaneous value from that recording site at that time, as well as to remove some high frequency noise. The data is thus effectively down-sampled to one sample per multiplexing cycle.
The data of the 16 down-sampled waveforms are then rearranged in a 16 x 16 x N matrix, or in a 64 x 4 x N matrix, where N is the number of samples (i.e., the time dimension) while the other two are the spatial dimensions of the recording matrix.

*Time-lag correction*
As described in the previous 'Data acquisition' section, the recording sites from the same row are accessed at different times, due to the sequential multiplexing approach employed, and thus slightly mismatched in time. If the first column is chosen as a reference, the mismatch increases from 1/16 of a sample for the second column, up to 15/16 of a sample for the 16th column.
This can be corrected by shifting in time the signals of each column by a proper amount in order to realign them to the one chosen as a reference, e.g., the first one.
Two simple approaches can thus be employed. The first approach uses the Fourier Shifting Theorem, in which case care must be taken to correctly implement a shift by a fraction of a sample. This approach has the advantage of requiring the same number of samples for the input and the output signal, i.e., it does not require additional memory.

Another approach is to first up-sample (i.e., interpolate) the signals by a factor of 16, which is performed by Fourier trasform, zero-padding and inverse Fourier transform. The interpolated signals are then shifted in time by discarding an integer number of samples at the beginning: 1 for the second column, 2 for the third and so on, up to 15 for the last column. In other words, the column-specific time shifts are proportional to the ordinal position of the column in the recording matrix.
In this approach, up-sampling requires 16 times the memory of the initial data, but is straightforward to implement, improves subsequent digital filtering performance and allows better visualization of the waveforms.

Both calibration and evoked signals were reconstructed and corrected for time-lags using the method described above.

*Calibration and Filtering*
Each recording site transistor has slightly different values of resistance and gain (due to the manufacturing process), resulting respectively in different values of DC offset and scaling factor in the recorded signals.
The offsets are cancelled by high-pass filtering the signals at a cutoff frequency (fs= 2 Hz) below the one of the amplifier. A low-pass filter was also used to remove high-frequency noise (fs= 300 Hz).
In order to compensate for the gain mismatch, a scaling factor is computed from a calibration sequence recorded just before or after each data acquisition. When recording a calibration signal, a sinusoidal current with known amplitude and frequency is injected by the reference electrode. The 16x16 scaling factors are then computed, independently for each recording site, from the amplitude of the largest Fourier component of the recordings. The calibration data are then rescaled pixelwise in such a way to display exactly the same amplitude of the sinusoidal waveform. The recorded neuronal signals are rescaled in the same way, using the same scaling factors from the respective calibration data. The final data is thus considered to be a measure of the Local Field Potentials at the recording sites.

Calibrated signals are spatially smoothed using a Gaussian filter kernel with one standard deviation. Smoothed signals were then visualized as 2D and 3D frames using custom Matlab scripts and time-lapse videos were generated from the single frames. In order to draw equipotential lines over the 2D frames, signals are spatially smoothed using a Gaussian filter kernel with two standard deviations.

############ Cortical Depth reconstruction

*Laminar profile creation (MUFTI)*
The processed signals acquired from each recording depth were concatenated to obtain a laminar profile of the evoked response across the cortical depth of the barrel column. This reconstruction was done under the assumption that the evoked response to a given set of repeated stimuli at a specific depth is reproducible during the whole experiment. However, as the signals were acquired at discrete times, mismatches in offset and peak amplitude of the evoked response were noticed in signals acquired from subsequent recording depths. For the correction of these mismatches, half of the sensor matrix (i.e., 8×16 electrodes each) was overlapped during data acquisition from two consecutive recording depths. The mismatch correction was performed in two steps - signal rescaling and signal averaging. As for the signal rescaling, the maximum response peak amplitude was determined in a signal pool containing data from L number of recording depths (i.e., L×16×16 electrodes). The median of the response peak amplitudes at that depth ($l \in \{1, ..., L\}$ with 16×16 electrodes) was considered as reference amplitude ($RA^l$).
Based on the location of the RA signal and under the above-mentioned assumption, signals from the recording depth containing the corresponding overlapped area are rescaled to the reference amplitude.
Under this assumption, the $RA^l$ was used in rescaling the signals from other recording depths in upward and downward cascading fashion. In other words, upward cascading was applied to rescale signals from depths $l-1$ to 1, the downward cascading rescaled signals from depths $l+1$ to $L$. Finally, the laminar profile was obtained by averaging the rescaled signals from electrodes belonging to the overlapped regions between $l$ (bottom 8×16 electrodes) and $l+1$ (top 8×16) recording depths. The laminar profile was lowpass filtered smoothed using two-pass 3-point hamming filter and visualized as time-lapsed 2D and 3D frames.

*Histology*
In order to verify the position of the probe within the barrel field and to examine any sign of injury produced in the tissue, standard histological assays were performed. At the end of the recording sessions, after a careful extraction of the needle from the cortex, rats were perfused through the left ventricle with Phosphate Buffered Saline (PBS, in mM: 137 NaCl, 2.7 KCl, 10 Na2HPO4, 2 KH2PO4; pH 7.4), followed by 4% neutral buffered Formalin solution (Sigma-Aldrich Co, Missouri, USA).

Brains were removed from the skull and immersed in fixative (4% Formalin solution) overnight. Fixed brains were then placed in sucrose solutions with increasing concentrations (12%, 18%, 30%; solution changes were made when brains sank). Sections through the somatosensory cortex were cut on a vibrating blade microtome (Leica VT1000 S, Leica Biosystems GmbH, Nussloch, Germany). Needle chip tracks along the entire cortical depth were visualized in 30 µm thick coronal slices of the right hemisphere. Neuronal cells were visualized in both types of cortical sections via Nissl staining with Cresyl Violet Acetate 0.1% (C5042, Sigma-Aldrich Co, Missouri, USA). Cortices were never flattened prior to cut them, not to distort the shape of the insertion trace. Slices were observed and images were acquired by bright field microscopy through a confocal laser-scanning microscope (Zeiss LSM 800; Carl Zeiss AG, Germany).

Inflammation and microglia activation was evaluated on cryostat-cut 20 µm thick coronal slices using 4,6-diamino-2-phenylindole (DAPI) and anti-ionized calcium binding adaptor molecule 1 (Iba1) immunohistochemistry. Procedures were explained in detail in (Sorrenti et al, 2018). Briefly, tissues were cut at -20 °C through a cryostat (Leica Microsystems, Wetzlar, Germany), mounted on SuperFrost glass slides (Fisher Scientific, Milan, Italy) and stored at -20 °C. After blocking nonspecific staining by incubating the slices with 5% normal goat serum and 0.1% Triton X-100 in PBS for 1 h at room temperature, sections were incubated with Iba1 primary antibody (1:800; Wako Chemicals Inc., Japan) for 2 h, followed by Alexa Fluor 488 fluorescent-conjugated secondary antibody (1:1000; Invitrogen) for 1 h in the blocking solution. Slides were carefully washed with PBS between subsequent steps. Nuclei were stained with DAPI (0.1 µg/ml). Slices were imaged using a fluorescence microscope (DMR, Leica Imaging Systems, 10x and 20x objectives) equipped with a digital camera (DC 200, Leica Imaging Systems).

Author Contributions
Claudia Cecchetto performed the experiments, curated the setting up of the experiments, performed part of the data analysis, curated the related mechanical devices and wrote the manuscript. Vincenzo Sorrenti performed the histological and immunochemical analysis and contributed to the writing of this manuscript. Mufti Mahmud and Enrico Chiarello performed the analysis of all the generated data from the probes and contributed to the writing of this manuscript. Marta Maschietto performed the experiments with 4x64 probes and contributed to the writing of this manuscript. Mathias Schulz and Roland Thewes designed the TU probes and helped in writing the related methods part of this manuscript. Stefano Vassanelli designed the project and related experiments and partially contributed to the writing of this manuscript.

Aknowledgement
We thank Sven Schroeder, Stefano Girardi, Roberto Oboe and Bernd Kuhn for useful discussions, scientific contributions to past experiments on this project and useful tips for the writing of this paper.
This project was funded by EU Projects: CYBERRAT, RAMP to Prof. Vassanelli and GRACE project to dr. Claudia Cecchetto.

**Figures**

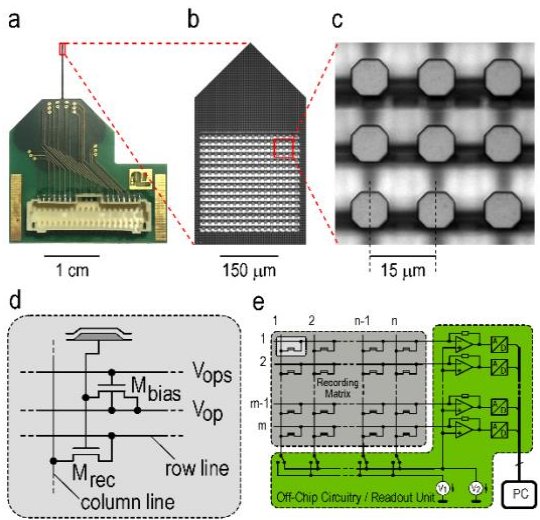

Figure 1: a) A fully assembled probe; b) magnification of the probe needle tip (16x16 array); c) magnification of a subset of recording sites on the chip surface (16x16 array); d) circuitry of a single recording site with recording transistor Mrec and biasing transistor Mbias; (e) simplified circuit diagram of a recording array with off-chip circuitry.

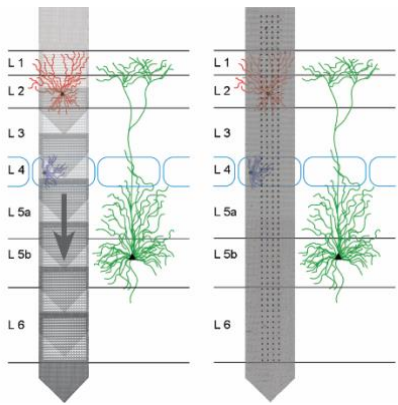

Figure 2: A magnification of the tip of the 16x16 array (left) and 4x64 array (right) chips with superimposed cortical neurons of the barrel cortex. The 16x16 array is moved by successive 120-µm vertical steps in order to sample the whole cortical column (as indicated by the arrow). A single insertion and no further movements are needed for recording from the whole cortical column with the 4x64 array chip.

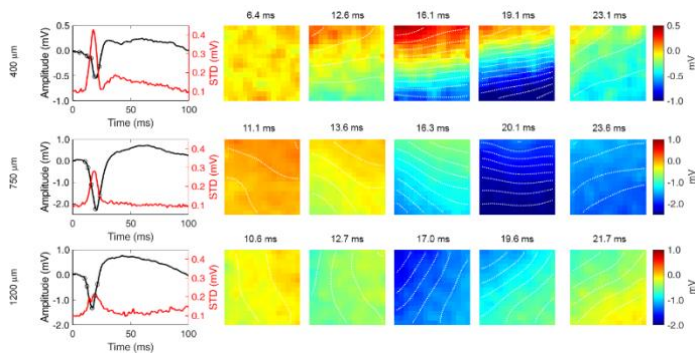

FIgure 3: Flow propagations of evoked LFPs recorded at different cortical depths by the 16x16 array. Single evoked responses are shown for layer III (400 mm, top), layer IV (750 mm, center) and layer Vb (1200 mm, bottom). For each depth, the time course of the amplitude (mV) and STD (mV) is shown on the left panel. The five panels on the right show the electrical potentials probed by the 256 sensors on the array tip at five distinctive time instants of the evoked response, color-mapped in a 16×16 image where each pixel is a single recording site. Dashed white lines plotted over the 16x16 panels are equipotential lines.

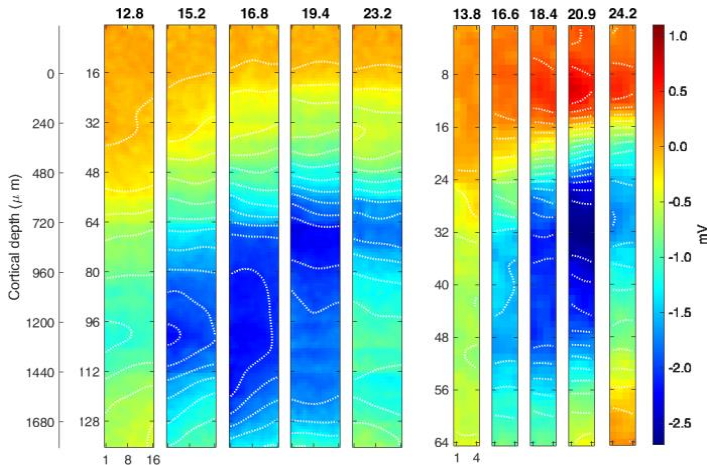

figure 4: Full cortical profile recorded by the 16x16 array (15 μm resolution, first 5 columns on the left) and the 4x64 array (32 μm resolution, 5 columns on the right) at five distinctive time instants of the evoked response. For the 16x16 array, the cortical profile across the whole depth of the barrel column is reconstructed by concatenating the 16×16 signals recorded from successive depths. For the 4x64 array, the cortical profile is recorded with a single insertion of the probe. The lateral extension of the recorded LFP profile is 225 μm for the 16x16 array and 96 μm for the 4x64 array.

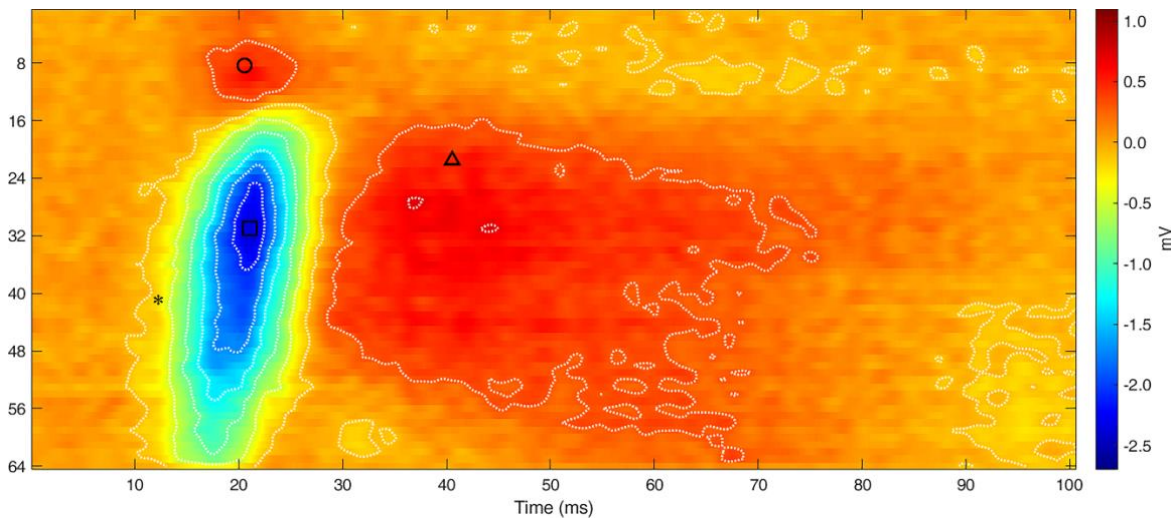

Figure 5: Time course of a single evoked response recorded by the 4x64 array. The LFP signal in each row is obtained by averaging along the 4 columns of the corresponding row at every time frame. The average signal is plotted on a 2D color map versus time, showing the first 100 ms of the signal following a single whisker deflection. Row 1 corresponds to the cortical surface (0 μm) and row 64 corresponds to about 2016 μm deep in the cortex.

| Chip # | Layout | Pitch (µm) | Total number of experiments | Total time of implantation |
|---|---|---|---|---|
| 1 | 16x16 | 15 | 2 | 8h |
| 2 | 16x16 | 15 | 13 | 42h 10min |
| 3 | 16x16 | 10 | 14 | 39h 30min |
| 4 | 4x64 | 32 | 2 | 6h |
| 5 | 4x64 | 32 | 20 | 43h 30min |

Table 1 Probe list with technical specifications and usage.

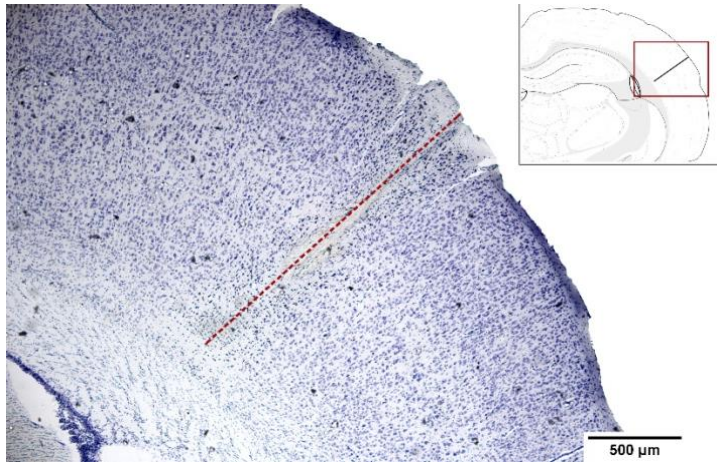

Figure S1: A 30-µm thick coronal slice in correspondence to the barrel field (in the inset, a schematic drawing of the anatomical atlas of the brain at AP -2.85). Neuronal cells are visualized in the cortical section via Nissl staining with Cresyl Violet Acetate. The orthogonality of the insertion and the correct position of the probe into the barrel field is clearly verified (the red dashed line denotes the trace of the insertion).

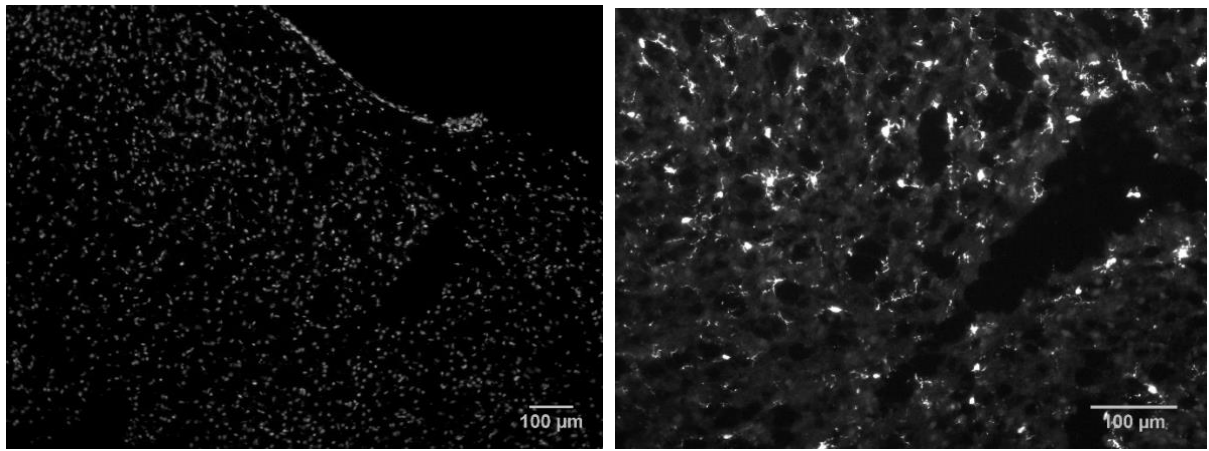

Figure S2: Inflammation and microglia activation around the probe insertion assessed on cryostat-cut 20 µm thick coronal slices using DAPI (left) and Iba1 (right) immunohistochemistry. The neuronal architecture (i.e. the position of somata) of the cortex after probe insertion is revealed by DAPI immunohistochemistry. Activated microglia appears as bright cells around the probe insertion and it is restricted to a small portion of tissue.